# scientific reports

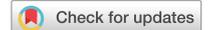

OPEN

# Total nitrogen estimation in agricultural soils via aerial multispectral imaging and LIBS

Md Abir Hossen[1], Prasoon K Diwakar[2] & Shankarachary Ragi[1]✉

Measuring soil health indicators (SHIs), particularly soil total nitrogen (TN), is an important and challenging task that affects farmers' decisions on timing, placement, and quantity of fertilizers applied in the farms. Most existing methods to measure SHIs are in-lab wet chemistry or spectroscopy-based methods, which require significant human input and effort, time-consuming, costly, and are low-throughput in nature. To address this challenge, we develop an artificial intelligence (AI)-driven near real-time unmanned aerial vehicle (UAV)-based multispectral sensing solution (UMS) to estimate soil TN in an agricultural farm. TN is an important macro-nutrient or SHI that directly affects the crop health. Accurate prediction of soil TN can significantly increase crop yield through informed decision making on the timing of seed planting, and fertilizer quantity and timing. The ground-truth data required to train the AI approaches is generated via laser-induced breakdown spectroscopy (LIBS), which can be readily used to characterize soil samples, providing rapid chemical analysis of the samples and their constituents (e.g., nitrogen, potassium, phosphorus, calcium). Although LIBS was previously applied for soil nutrient detection, there is no existing study on the integration of LIBS with UAV multispectral imaging and AI. We train two machine learning (ML) models including multi-layer perceptron regression and support vector regression to predict the soil nitrogen using a suite of data classes including multispectral characteristics of the soil and crops in red (R), near-infrared, and green (G) spectral bands, computed vegetation indices (NDVI), and environmental variables including air temperature and relative humidity (RH). To generate the ground-truth data or the training data for the machine learning models, we determine the N spectrum of the soil samples (collected from a farm) using LIBS and develop a calibration model using the correlation between actual TN of the soil samples and the maximum intensity of N spectrum. In addition, we extract the features from the multispectral images captured while the UAV follows an autonomous flight plan, at different growth stages of the crops. The ML model's performance is tested on a fixed configuration space for the hyper-parameters using various hyper-parameter optimization techniques at three different wavelengths of the N spectrum.

Soil health indicators are a composite set of measurable physical, chemical and biological properties which can be used to determine soil health status. Among the chemical indicators, we particularly focus on nitrogen (N) because N is the most limiting nutrient in many of the world's agricultural areas[1]. Insufficient use of N causes economic loss, in contrast, excessive use of N implies wasting fertilizer, causes nitrate pollution, and increases the cost[2,3]. Nitrogen treatment can account for up to 30% of the total production cost[4].

Chlorophyll meter (CM) measures the chlorophyll content of crops to estimate their N nutrition status. In recent years, the use of CM has increased among researchers and farmers[5,6]. For instance, N application rates for corn were determined using the adjusted $R^2$ of the relationship between nitrogen rate difference (ND) and CM readings[7]. However, CM-based methods fail to capture the spatial variability that is often present within the field. For N management, determination of spatial patterns is necessary but requires collection and analysis of a large number of samples which is labor-intensive and time-consuming[2,6].

Satellite-based remote sensing is one alternative to ground-based measurements. Satellite-based techniques utilize images at the spectral level for crop growth monitoring and real-time management[8–10]. For instance, vegetation indices (VIs), evaluated using the data obtained from satellite-based multispectral sensors, have been used to detect the N stress at V4–V7 (4–7 leaves with visible leaf collar) stages[11–13]. However, satellite-based sensing suffers from lower spatial and temporal resolution, and sensing disruption may occur during image acquisition in some areas because of cloud cover and/or sprinkler irrigation[14]. Farmers' adaption of the system is still limited. Additionally, the high cost of obtaining these images for relatively small areas is a significant drawback[15]. Multispectral cameras mounted on unmanned aerial vehicles (UAVs) have enormous potential to resolve this problem. UAVs can be deployed rapidly and frequently for image acquisition, resulting in reduced costs, greater flexibility in terms of data resolution and mission timing[16–18]. For instance, a variable rate N fertilization map was created using hyperspectral airborne images[19], the ground-sensor measurements were compared with hyperspectral images to determine the N sufficiency index[17], and estimation of N side-dress using NDVI computed from aerial imaging[20]. However, radiometric and geometric calibrations are needed for the UAVs on-board miniaturized electro-optical sensors to obtain quantitative results and provide precise georeferencing[21]. UAVs also fail to perform on-board mosaicking images due to limited computational resources.

[1]Department of Electrical Engineering, South Dakota School of Mines and Technology, Rapid City, SD 57701, USA. [2]Department of Mechanical Engineering, South Dakota School of Mines and Technology, Rapid City, SD 57701, USA. ✉email: Shankarachary.Ragi@sdsmt.edu

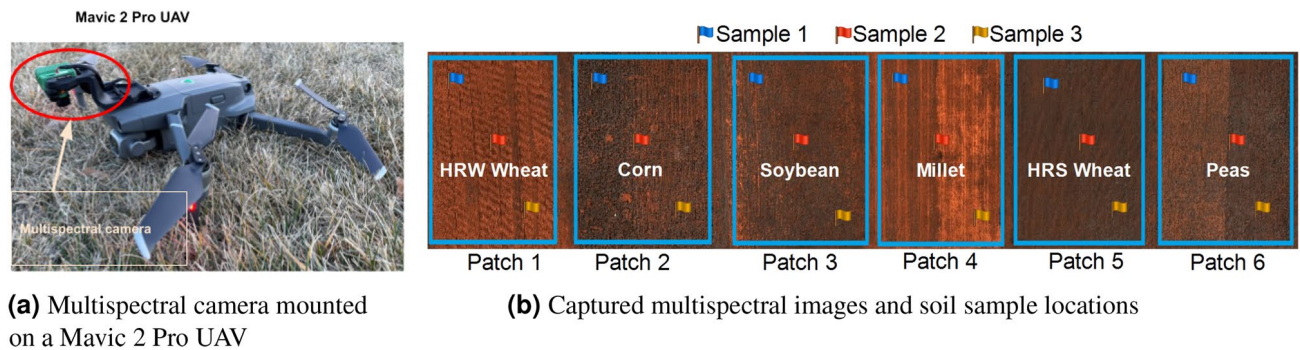

**Figure 1.** (a) Mavic 2 pro UAV with multispectral camera mounted. (b) The flags show the sample locations of the corresponding crops. The patches have crops including Peas, HRS Wheat, Millet, Soybean, Corn, and HRW Wheat, respectively.

Laser induced breakdown spectroscopy (LIBS) is an analytical method for qualitative and quantitative elemental detection. LIBS can be readily applied to soil samples, providing rapid chemical analysis of soil samples and their constituents (e.g., Nitrogen, Potassium, Phosphorus, Calcium). The combination of an autonomous UA, LIBS, and machine learning can be used to achieve in-field measurement which provides instant results for deficient nutrient analysis and fertilization planning. With appropriate calibration, the LIBS analysis can provide quantitative measurement for most elements in soil including, carbon, nitrogen, potassium, sulfur, and phosphorus[22,23]. There have been some applications of standalone LIBS systems in precision agriculture[22–25]. However, there has been no detailed research of LIBS application in combination with ML and UAVs. Some studies have found it challenging to measure nitrogen using LIBS due to environmental factors; Earth's atmosphere is almost 80% nitrogen which will interfere with the sample measurement result since the soil is less than 1% nitrogen. Testing in a vacuum or in low-pressure conditions has been suggested to improve measurement accuracy[23]. In this study, we conducted LIBS analysis on soil samples under a normal atmosphere for observation. Low laser pulse energies were used to minimize the breakdown of air and thereby minimize the influence of atmospheric nitrogen.

The purpose of the present study is to develop a machine learning (ML)-based predictive model to estimate TN of soil using crops and soil spectral characteristics measured from the multispectral images captured from a UAV, and LIBS. Specifically, we train a multi-layer perceptron regression (MLP-R) and support vector regression (SVR) model to predict TN in soil. We use root mean square error (RMSE) and computational time (CT) as performance metrics to measure the performance of the above predictive model. To reduce the RMSE and lower CT in the machine learning models, we perform hyper-parameter optimization (HPO). The HPO tuning process depends on the ML model used for prediction[26]. The traditional way to tune hyper-parameter is through manual testing, although it requires a deep understanding of the ML models[27]. However, manual tuning is ineffective for many problems due to a large number of hyperparameters, model complexity, time-consuming model evaluations, and non-linear hyper-parameter interactions. Several HPO techniques[28] have been used for different applications such as grid search (GS), random search (RS), bayesian optimization, genetic algorithm (GA), and particle swarm optimization. In this study, we implement GS, RS, and GA for hyper-parameter optimization.

### Experimental design

An aerial survey was carried out with Mavic 2 Pro UAV. We obtained multispectral images using the *Sentera* high-precision NDVI single sensor which was mounted on the UAV (Fig. 1a). The sensor is 1.2 MP CMOS with a $60°$ horizontal FOV and a $47°$ vertical FOV and works with two wide spectral bands: red (625 nm CWL × 100 nm width) and NIR (850 nm CWL × 40 nm width) with a pixel count of 1248 horizontal/950 vertical. The green band is typically unused. The sensor has a total weight of 30 g and size of 25.4 × 33.8 × 37.3 mm.

**Data collection: multispectral images and soil samples.** The farm used for data collection is located at Sturgis, South Dakota, USA ($44°\ 25'\ 27''N$; $103°\ 22'\ 34''W$). We created an autonomous UAV flight plan for minimal passes similar to a raster scan pattern using the coordinates of the four corners of the field (44.25.39 N, 103.22.60 W; 44.25.28 N, 103.22.60 W; 44.25.39 N, 103.23.16 W; 44.25.39 N, 103.23.16 W). We captured 865

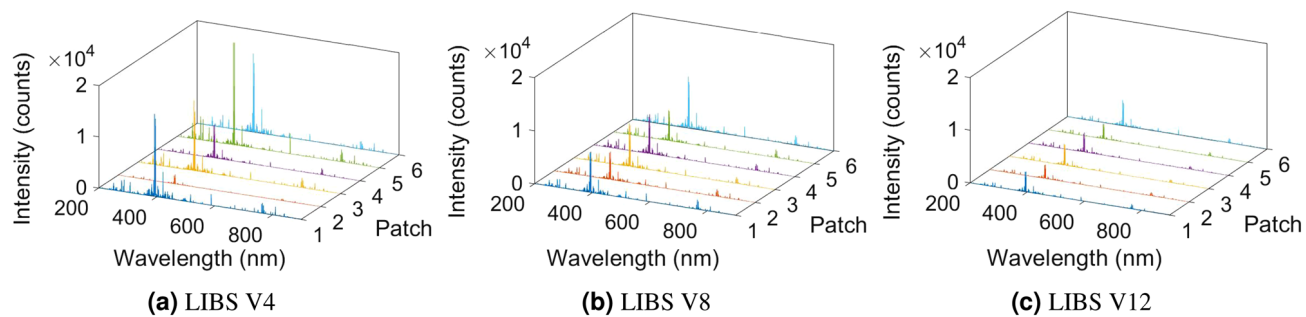

**Figure 2.** Emission lines of soil samples at the V4, V8, and V12 stages for six patches.

multispectral images at each of the growth stages (V4, V8, and V12) and used *Sentera* image stitching software to mosaicking the images. The multispectral images were captured while the UAV was following the raster scan pattern using the following parameters and experimental setup,

(i) Parameters:

- Flight Type: *QuickTile*
- Overlap Setting: 75%
- Altitude: 60.96 m
- Speed: 6.71 m/s

(ii) Experimental setup:

- Desired resolution: The Ground Sample Distance (GSD)/pixel of the multispectral camera was set to 0.05 m for a 60.96 m altitude.
- Cloud cover and time of day: The UAV was flown when the sun was highest in the sky for more accurate data. Data is best when sky conditions are consistent, ideally 100% sunny or 100% cloudy. Flying with a mix of sun and clouds causes inconsistency in brightness and contrast while stitching images. Therefore, the stitched image will provide an inaccurate NDVI value.

We collected six soil samples 6.1 m from the edge of the field and six samples from the opposite side of the field, and six soil samples from the center of the field as shown in Fig. 1b. We followed the soil sampling methods for South Dakota region[29] to select the sample locations and number of samples collected. A total of 54 soil samples were collected at an 0.2 m depth from six patches (3 samples per patch) at the V4, V8, and V12 stages (18 samples per stage) using a hydraulic probe. We avoided sampling from the areas where conditions were different from the rest of the field (e.g., former manure piles, fertilizer bands, or fence lines). Figure 1b shows the sample locations across the patches.

**Calibration.** LIBS utilizes a high energy pulsed laser which generates a high temperature ranging from 10°–20,000°K resulting in plasma formation when focused on a sample. This, in turn, leads to ablation of a minuscule amount of sample, leading to excitation of the sample's constituent elements. As the plasma cools, these excited atoms and electrons emit photons which correspond to specific elements present in the sample. These photons are collected by a spectrometer and result in quantitative and qualitative analysis of samples. The SciAps Z-300 handheld LIBS analyzer was used for these measurements. This device has an extended spectrometer wavelength range from 190 $\mu$m to 950 $\mu$m. The extended range allows emission lines from elements H, F, N, O, Br, Cl, Rb and, S to be measured. The LIBS instrument is equipped with a Q-switched Nd:YAG laser, 5-6 mJ per pulse at 1064 nm. Ten laser pulses are shot on the soil samples in the presence of Ar purge to obtain averaged data on each measurement. The focused laser on the soil surface forms a $\mu m$ size of a sample into $> 10,000°$K plasma. The unique emission spectrum is collected by the spectrometer as the plasma cools.

We used NIST LIBS database[30] to determine the N lines from the emission spectrum (Fig. 2) and found N lines at 493.4 nm, 746.6 nm, 821.4 nm, and 868.1 nm (Fig. 3). However, we discarded the 746.6 nm N lines due to weaker intensity response and inconsistency between the samples in wavelength. We verified the N lines from the study of soil nutrient detection for precision agriculture[22]. From the soil samples, we select four samples randomly and obtained the actual TN of soil in ppm for calibration. We analyzed all the 54 soil samples in LIBS to determine the N spectrum's maximum intensity at 493.4 nm, 821.4 nm, and 868.1 nm (Figs. 5, 6, and 7) at V4, V8, and V12 stages.

Using the correlation between actual TN and the maximum intensity of N spectrum, we construct calibration plots for 493.4 nm, 821.4 nm, and 868.1 nm through linear regression (Fig. 4). We use $R^2$ as our calibration metric and find $R^2 = 0.98$, $R^2 = 0.99$, and $R^2 = 0.90$, respectively, showing a strong correlation between the actual soil TN and the peak intensity of the N spectrum. Using the calibrated model, we converted the peak intensity of the N spectrum (Figs. 5, 6, and 7) to TN (ppm) for all the 54 soil samples (Table 1) to generate the training data for the ML models.

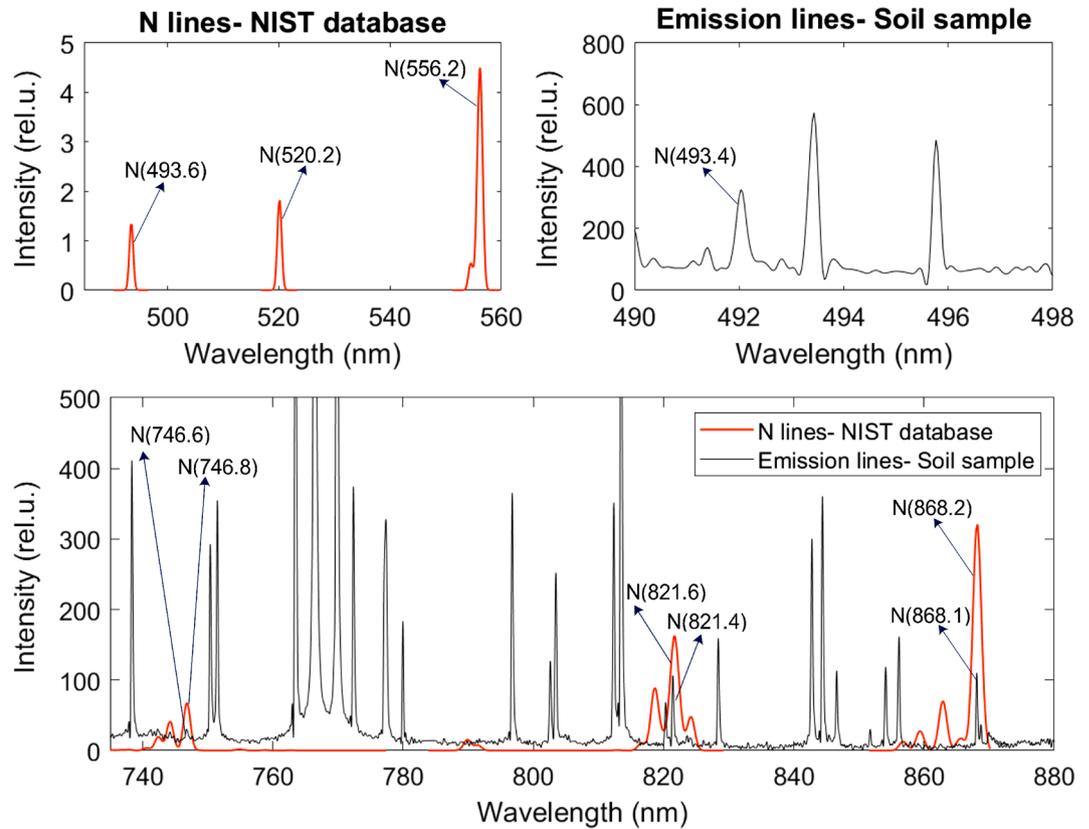

**Figure 3.** Determining N lines from the soil sample using NIST database.

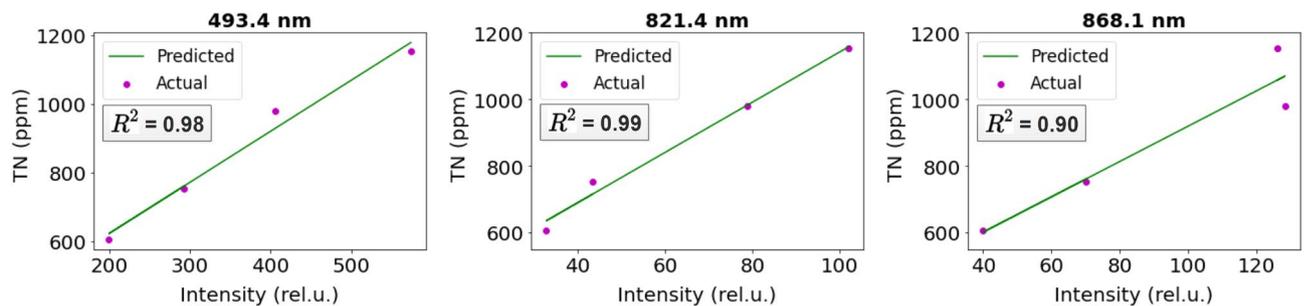

**Figure 4.** Calibration plot for computing soil TN using the peak intensity of the nitrogen spectrum at 493.4 nm, 821.4 nm, and 868.1 nm.

**Feature extraction and dataset.** The multispectral images are composed of three channels, channel-1: R, channel-2: G, and channel-3: NIR. The multispectral sensor's datasheet[31] shows that channel-1 contains both R and NIR light. Therefore, the NIR light needed to be removed to isolate R and compute NDVI. The equations for R and NIR light are,

$$R = 1.0 * DN_{ch1} - 1.012 * DN_{ch3} \qquad (1)$$

$$NIR = 9.605 * DN_{ch3} - 0.6182 * DN_{ch1} \qquad (2)$$

where $DN_{ch1}$ is the Digital Number (pixel value) of channel one, and $DN_{ch3}$ is the Digital Number (pixel value) of channel three. The coefficients of $DN$ were provided in the datasheet[31].

Using Eqs. (1) and (2), band separation (Fig. 8a) was performed to compute NDVI (Fig. 8b) and extract the pixel values from each of the bands. The dataset (Table 1) was created using the mean NDVI and the mean pixel values of each of the bands from individual zones at the V4, V8, and V12 stages. The equation for computing NDVI,

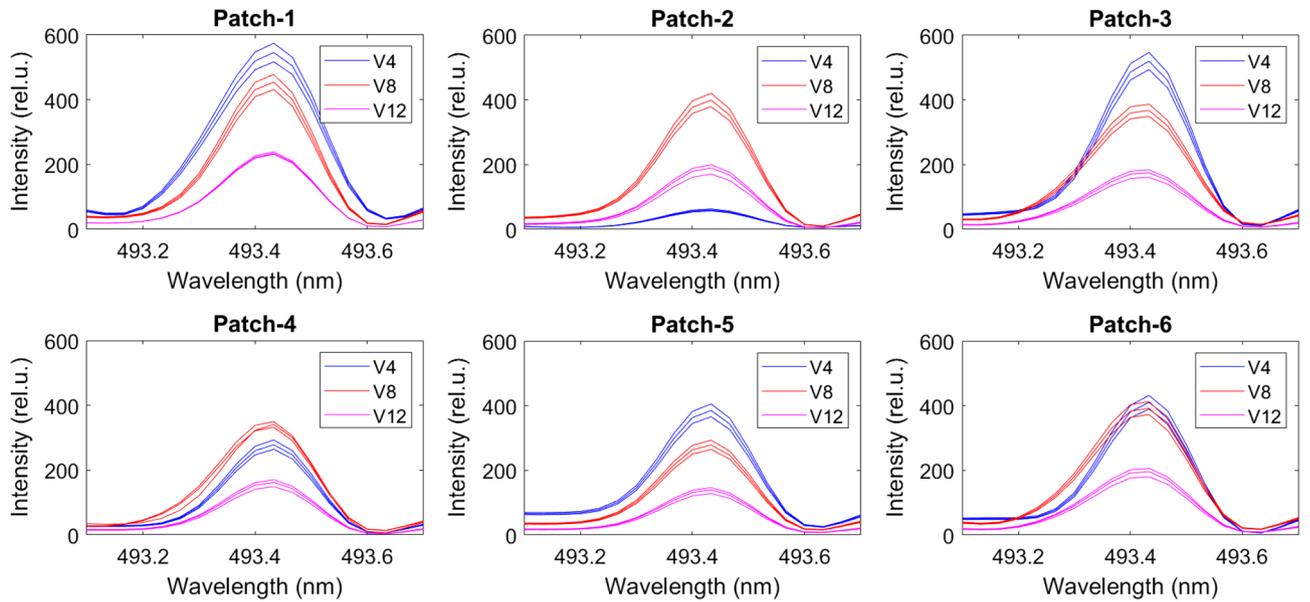

**Figure 5.** Nitrogen spectrum of the soil samples at 493.4 nm for six patches at the V4, V8 and V12 stages.

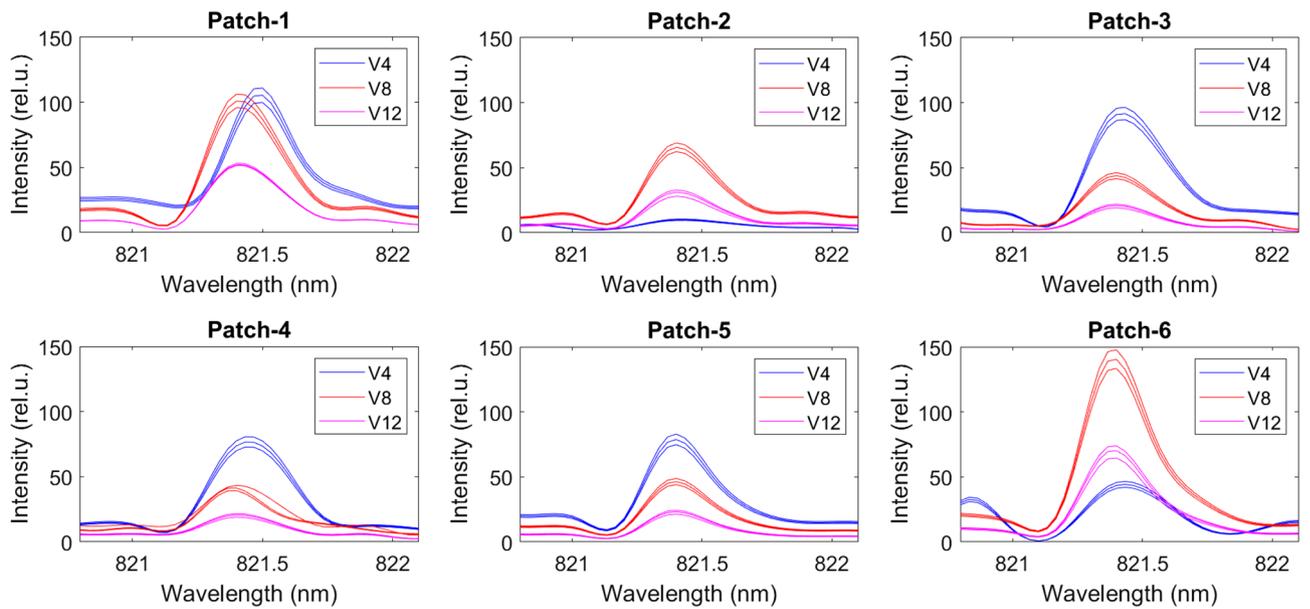

**Figure 6.** Nitrogen spectrum of the soil samples at 821.4 nm for six patches at the V4, V8 and V12 stages.

$$NDVI = \frac{1.236 * DN_{ch3} - 0.188 * DN_{ch1}}{1.000 * DN_{ch3} + 0.044 * DN_{ch1}} \qquad (3)$$

## Methods

In supervised learning, the goal is to obtain an optimal predictive model function $f^*$ based on the input $x$ and the output $y$ to minimize the cost function $L(f(x), y)$. In this study, we particularly use MLP-R and SVR which can be used for both classification and regression problems. We applied HPO techniques to determine the best set of hyper-parameters from the ML models and train the ML models using those hyper-parameters on the training dataset.

**Multi-layer perceptron regression (MLP-R).** Mulit-layer perceptron is a supervised learning algorithm that learns a function $f(.) : R^x \rightarrow R^o$ by training on a dataset[32], where $x$ is the number of input dimension and $o$ is the number of output dimension. We designed the MLP-R (Fig. 9a) with multiple organized layers consisting of various neuron-like processing units. Each node in the layer was connected with the nodes in the previous layer. Each node may have symmetrical or differing strengths and weights. The data in the network enters with

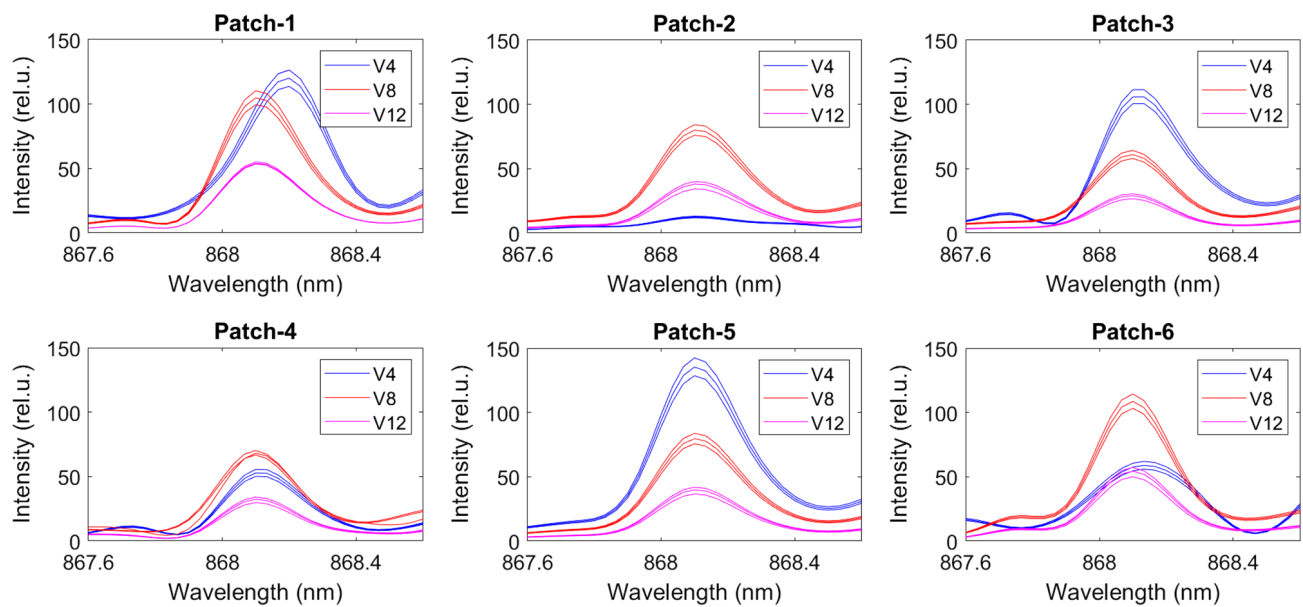

**Figure 7.** Nitrogen spectrum of the soil samples at 868.1 nm for six patches at the V4, V8 and V12 stages.

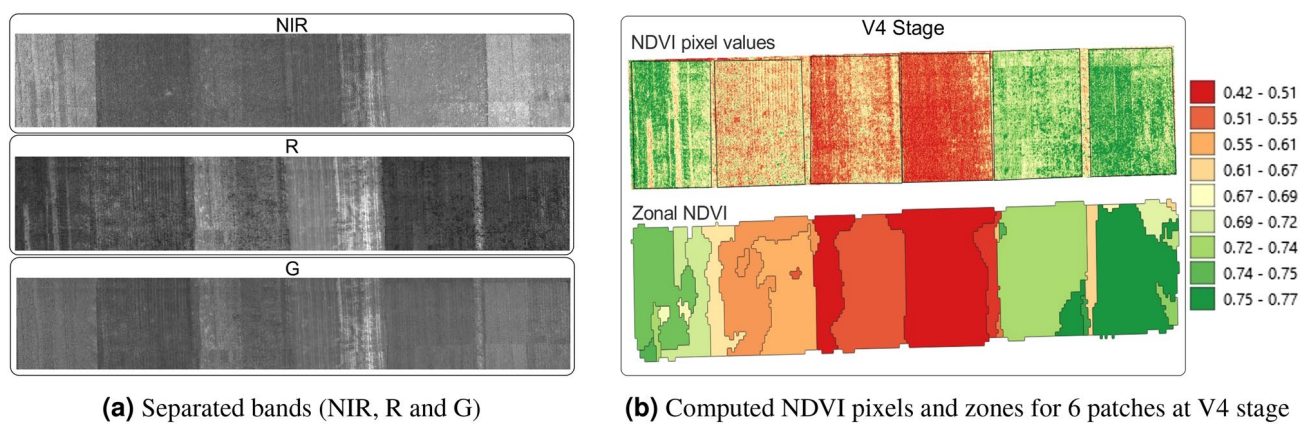

(**a**) Separated bands (NIR, R and G)  (**b**) Computed NDVI pixels and zones for 6 patches at V4 stage the input layer and gradually runs through each layer to reach the output layer. For a given a set of features $x = \{R, NIR, G, NDVI, \text{Air temperature}, RH\}$ and target $y = TN$, $f(.) : R^6 \rightarrow R^1$. To train the MLP-R from a given set of input-output pairs $X = \{(\vec{x}_1, y_1), \ldots, (\vec{x}_N, y_N)\}$, learning consists of iteratively updating the values of weight and bias of the perceptron to minimize RMSE. The hyper-parameter configuration (Table 2) was created using the solver type[33], activation function[34], learning rate and hidden layer sizes.

**Figure 8.** Band separation, and computed NDVI pixels and zonal NDVI.

**Support vector regression (SVR).** Support vector machine (SVM) makes data points linearly separable by mapping them from low-dimensional to high-dimensional space. The classification boundary creates a partition between the data points by generating a hyperplane[35]. SVM concepts can be applied to regression problems by generalizing them. SVR uses a symmetrical loss function that penalizes both high and low mis-estimates equally. The $\varepsilon$-tube is used to generalize SVM to SVR by adding an $\varepsilon$-insensitive region around the function, ignoring the absolute values of errors less than a certain threshold $\varepsilon$ from both above and below the estimation[36,37]. In SVR, points outside the tube are penalized, but points inside the tube, whether above or below the function, are not penalized. SVR uses different types of kernels for non-linear functions to map the data into a higher dimensional space[34,36,37]. Linear kernels, radial basis function (RBF), polynomial kernels, and sigmoid kernels are common kernel types in SVR. We created the hyper-parameter configuration (Table 2) using the kernel types, regularization parameter ($C$)[34], and distance error (*epsilon*) of the loss function[34].

**Hyper-parameter optimization (HPO).** GS, RS, and GA HPO techniques were executed within their respective hyper-parameters to train the model. We performed cross-validation by splitting the train and test data into 5-folds. After obtaining the RMSE from the cross-validation score, we selected the hyper-parameters which yielded the lowest RMSE. Finally, using the best set of hyper-parameters we trained the MLP-R and SVR

| Data count | Red (DN) | NIR (DN) | Green (DN) | NDVI | RH (%) | Air temp (C) | TN (ppm) at 493.4 nm | TN (ppm) at 821.4 nm | TN (ppm) at 868.1 nm |
|---|---|---|---|---|---|---|---|---|---|
| 0 | 47.88 | 328.05 | 54.90 | 0.74 | 33.8 | 23.2 | 1179.39 | 1156.89 | 1057.54 |
| 1 | 51.99 | 329.97 | 56.50 | 0.72 | 33.8 | 23.2 | 1136.64 | 1118.38 | 962.57 |
| 2 | 52.37 | 325.90 | 56.14 | 0.72 | 33.8 | 23.2 | 1093.90 | 1079.95 | 933.92 |
| 3 | 55.81 | 200.32 | 43.11 | 0.56 | 33.8 | 23.2 | 418.46 | 466.82 | 458.29 |
| 4 | 59.30 | 209.11 | 45.47 | 0.55 | 33.8 | 23.2 | 413.75 | 462.82 | 454.84 |
| 5 | 56.36 | 208.75 | 44.72 | 0.57 | 33.8 | 23.2 | 409.28 | 459.04 | 451.55 |
| 6 | 74.64 | 223.63 | 52.27 | 0.50 | 33.8 | 23.2 | 1139.47 | 1110.15 | 980.61 |
| 7 | 78.95 | 229.52 | 54.50 | 0.49 | 33.8 | 23.2 | 1098.66 | 1073.99 | 950.90 |
| 8 | 79.21 | 230.47 | 55.09 | 0.49 | 33.8 | 23.2 | 1059.94 | 1039.71 | 922.78 |
| 9 | 85.53 | 226.03 | 55.86 | 0.44 | 33.8 | 23.2 | 760.72 | 996.15 | 683.98 |
| 10 | 98.47 | 238.19 | 61.73 | 0.41 | 33.8 | 23.2 | 738.83 | 965.72 | 669.23 |
| 11 | 102.15 | 247.70 | 64.07 | 0.41 | 33.8 | 23.2 | 718.12 | 936.81 | 655.23 |
| 12 | 44.84 | 287.68 | 49.64 | 0.73 | 33.8 | 23.2 | 927.98 | 1013.97 | 1086.19 |
| 13 | 47.37 | 312.47 | 53.44 | 0.74 | 33.8 | 23.2 | 897.75 | 982.64 | 1069.74 |
| 14 | 48.64 | 323.58 | 55.13 | 0.74 | 33.8 | 23.2 | 869.15 | 952.89 | 1071.34 |
| 15 | 47.84 | 339.04 | 56.60 | 0.75 | 33.8 | 23.2 | 967.15 | 739.38 | 717.83 |
| 16 | 47.60 | 381.57 | 61.43 | 0.78 | 33.8 | 23.2 | 938.85 | 721.78 | 701.39 |
| 17 | 48.77 | 391.88 | 62.88 | 0.78 | 33.8 | 23.2 | 908.17 | 705.10 | 685.73 |
| 18 | 99.79 | 256.35 | 58.14 | 0.49 | 46.2 | 24.1 | 1036.11 | 1190.11 | 974.25 |
| 19 | 104.66 | 265.32 | 59.71 | 0.49 | 46.2 | 24.1 | 1000.51 | 1150.09 | 945.07 |
| 20 | 104.12 | 268.98 | 63.05 | 0.46 | 46.2 | 24.1 | 966.70 | 1112.19 | 917.05 |
| 21 | 46.67 | 311.63 | 52.52 | 0.72 | 46.2 | 24.1 | 950.92 | 909.48 | 839.01 |
| 22 | 43.28 | 358.41 | 57.92 | 0.75 | 46.2 | 24.1 | 919.64 | 883.35 | 813.49 |
| 23 | 41.69 | 354.49 | 56.49 | 0.76 | 46.2 | 24.1 | 889.85 | 858.59 | 792.27 |
| 24 | 65.41 | 308.62 | 57.18 | 0.67 | 46.2 | 24.1 | 901.02 | 736.13 | 729.72 |
| 25 | 70.86 | 338.76 | 59.38 | 0.70 | 46.2 | 24.1 | 872.28 | 718.69 | 712.69 |
| 26 | 78.54 | 353.09 | 59.30 | 0.74 | 46.2 | 24.1 | 844.87 | 702.15 | 696.50 |
| 27 | 39.49 | 260.42 | 63.32 | 0.46 | 46.2 | 24.1 | 845.17 | 714.99 | 761.44 |
| 28 | 51.39 | 264.63 | 66.37 | 0.44 | 46.2 | 24.1 | 831.91 | 699.59 | 750.14 |
| 29 | 56.50 | 269.17 | 66.83 | 0.45 | 46.2 | 24.1 | 819.11 | 684.03 | 742.82 |
| 30 | 55.99 | 311.83 | 51.53 | 0.76 | 46.2 | 24.1 | 761.02 | 755.98 | 833.60 |
| 31 | 63.82 | 325.55 | 51.81 | 0.79 | 46.2 | 24.1 | 739.13 | 737.56 | 811.37 |
| 32 | 66.80 | 336.38 | 54.13 | 0.78 | 46.2 | 24.1 | 718.42 | 720.12 | 790.25 |
| 33 | 48.12 | 349.84 | 59.99 | 0.73 | 46.2 | 24.1 | 938.85 | 1503.43 | 996.00 |
| 34 | 51.06 | 387.93 | 62.34 | 0.77 | 46.2 | 24.1 | 908.17 | 1447.56 | 965.76 |
| 35 | 49.63 | 400.78 | 63.20 | 0.78 | 46.2 | 24.1 | 878.98 | 1394.71 | 936.58 |
| 36 | 99.95 | 277.48 | 66.28 | 0.46 | 57.6 | 24.8 | 680.29 | 788.98 | 681.60 |
| 37 | 104.13 | 271.61 | 67.01 | 0.44 | 57.6 | 24.8 | 673.15 | 780.98 | 675.71 |
| 38 | 102.69 | 263.35 | 65.67 | 0.43 | 57.6 | 24.8 | 669.57 | 777.05 | 672.84 |
| 39 | 44.99 | 318.10 | 52.71 | 0.73 | 57.6 | 24.8 | 621.91 | 635.49 | 601.27 |
| 40 | 41.99 | 318.73 | 51.85 | 0.75 | 57.6 | 24.8 | 607.02 | 623.03 | 590.66 |
| 41 | 40.22 | 294.12 | 48.77 | 0.75 | 57.6 | 24.8 | 578.87 | 599.47 | 570.50 |
| 42 | 66.78 | 265.91 | 54.33 | 0.58 | 57.6 | 24.8 | 598.23 | 553.12 | 550.87 |
| 43 | 76.19 | 305.64 | 61.84 | 0.59 | 57.6 | 24.8 | 584.53 | 544.81 | 542.75 |
| 44 | 78.82 | 315.68 | 63.89 | 0.59 | 57.6 | 24.8 | 563.82 | 532.28 | 530.50 |
| 45 | 39.91 | 301.39 | 49.75 | 0.76 | 57.6 | 24.8 | 578.12 | 546.10 | 569.60 |
| 46 | 57.28 | 320.28 | 57.90 | 0.69 | 57.6 | 24.8 | 565.46 | 539.00 | 560.58 |
| 47 | 56.41 | 325.96 | 58.54 | 0.70 | 57.6 | 24.8 | 546.10 | 530.62 | 546.84 |
| 48 | 57.30 | 227.48 | 46.38 | 0.59 | 57.6 | 24.8 | 542.67 | 571.77 | 611.35 |
| 49 | 66.44 | 256.15 | 52.73 | 0.59 | 57.6 | 24.8 | 531.80 | 562.55 | 600.21 |
| 50 | 66.61 | 254.53 | 52.76 | 0.58 | 57.6 | 24.8 | 515.12 | 548.59 | 583.34 |
| 51 | 48.11 | 233.98 | 44.48 | 0.65 | 57.6 | 24.8 | 631.59 | 939.75 | 692.53 |
| 52 | 50.50 | 258.24 | 47.95 | 0.66 | 57.6 | 24.8 | 616.25 | 917.71 | 677.35 |
| 53 | 49.29 | 273.04 | 49.43 | 0.68 | 57.6 | 24.8 | 592.87 | 875.35 | 654.27 |

**Table 1.** Generated training data for the ML models.

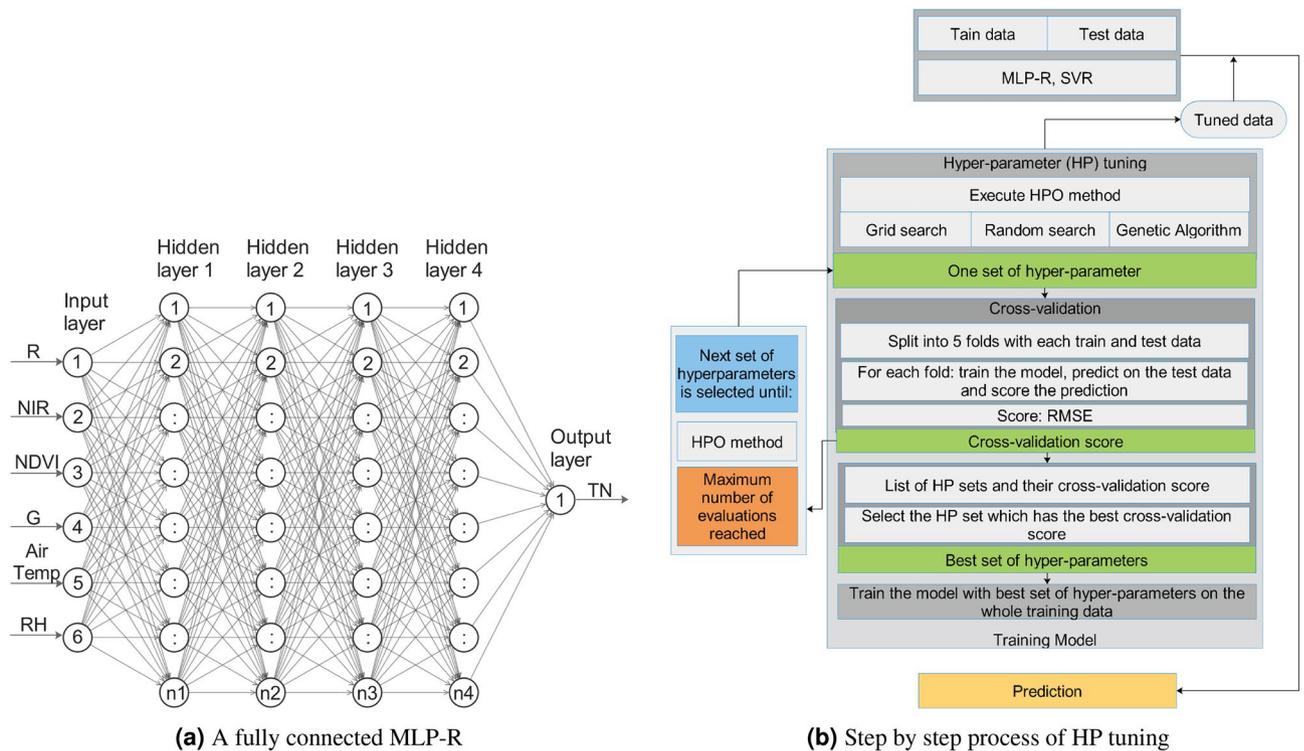

**(a)** A fully connected MLP-R  **(b)** Step by step process of HP tuning models for each HPO technique. Figure 9b shows the step-by-step process of HPO, training the dataset, and prediction of test data.

**Figure 9.** (a) MLP-R with four hidden network having different weights, where input layer $\in \mathbb{R}^6$, hidden layer $\in \mathbb{R}^4$, output layer $\in \mathbb{R}^1$ and n1, n2, n3, and n4 represent the number of perceptron in each hidden layer, respectively. (b) HPO for GS, RS and GA with cross-validation, training the ML models with the tuned HP and prediction.

| ML model | Hyper-parameter | Type | Search space | Default |
|---|---|---|---|---|
| MLP-R | activation | Categorical | ['relu', 'tanh', 'logistic', 'identity'] | 'relu' |
| | solver | Categorical | ['adam', 'lbfgs'] | 'adam' |
| | lerning_rate | Categorical | ['constant', 'adaptive', 'invscaling'] | 'constant' |
| | hidden_layer_sizes | Discrete | [20, 250] | 100 |
| SVR | kernel | Categorical | ['poly', 'rbf', 'sigmoid'] | 'rbf' |
| | C | Discrete | [1, 10000] | 1 |
| | epsilon | Discrete | [0.0001, 1] | 0.1 |

**Table 2.** Specifics of the configuration space for the hyper-parameters.

GS exhaustively evaluates all the combinations in the hyper-parameter configuration space specified by the user in the form of a grid configuration[38]. The user must identify the global optimums manually since GS cannot utilize the well-performing regions[28]. However, in RS, the user defines a budget (i.e., time) as well as the upper and lower bounds of the hyper-parameter values. RS randomly selects the values from the pre-defined boundary and trains until the budget is exhausted[28]. If the configuration space is wide enough, RS can detect the global optima. Assuming a model has $k$ parameters and each of them has $n$ distinct values, the GS computational complexity increases exponentially at a rate of $O(n^k)$[39]. Therefore, the effectiveness of GS depends on the size of the hyper-parameter configuration space. For RS, the computational complexity is defined as $O(n)$, where $n$ is specified by the user before the optimization process starts[28].

GA[40] randomly initializes the population and chromosomes. Genes represents the entire search space, hyper-parameters, and hyper-parameter values. GA uses a fitness function to evaluate the performance of each individual in the current generation similarly to the objective function of a ML model. To produce a new generation, GA performs selection, crossover, and mutation operations on the chromosomes involving the next hyper-parameter configurations to be evaluated. The cycle continues until the algorithm reaches the global optimum.

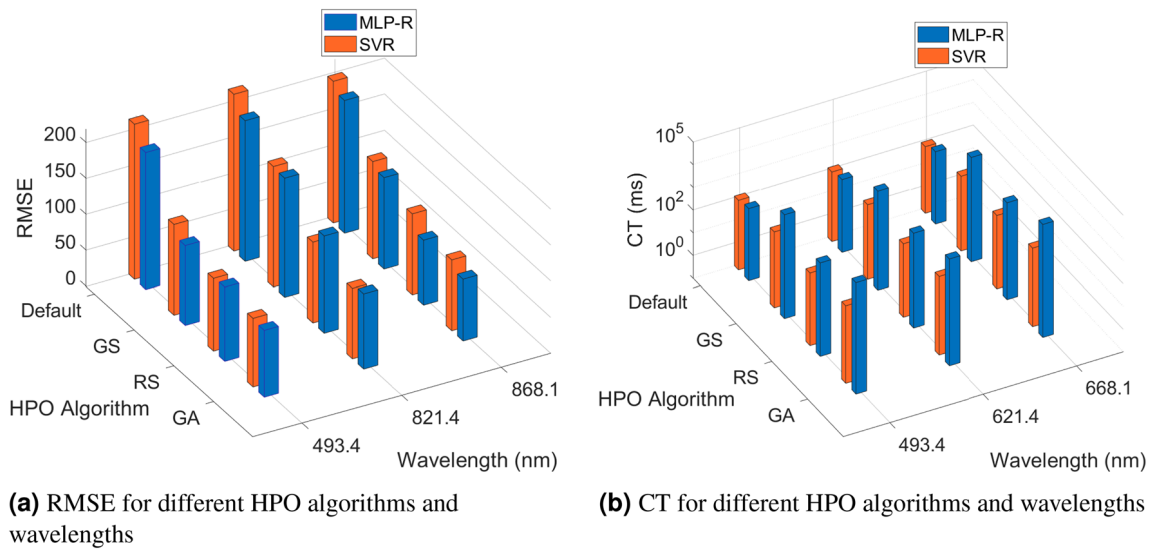

(a) RMSE for different HPO algorithms and wavelengths

(b) CT for different HPO algorithms and wavelengths

**Figure 10.** Performance comparison for different HPO algorithms at different wavelengths.

| HPO algorithm | ML model | Hyper-parameter | Best parameter for 493.4 nm | Best parameter for 821.4 nm | Best parameter for 868.1 nm |
|---|---|---|---|---|---|
| GS | MLP-R | activation | 'logistic' | 'logistic' | 'tanh' |
| | | solver | 'lbfgs' | 'lbfgs' | 'lbfgs' |
| | | learning_rate | 'constant' | 'constant' | 'constant' |
| | | hidden_layer_sizes | 50 | 50 | 250 |
| | SVR | kernel | 'poly' | 'rbf' | 'poly' |
| | | C | 1000 | 1000 | 1000 |
| | | epsilon | 1 | 1 | 0.01 |
| RS | MLP-R | activation | 'logistic' | 'logistic' | 'relu' |
| | | solver | 'lbfgs' | 'lbfgs' | 'lbfgs' |
| | | learning_rate | 'adaptive' | 'constant' | 'adaptive' |
| | | hidden_layer_sizes | 76 | 41 | 61 |
| | SVR | kernel | 'poly' | 'poly' | 'poly' |
| | | C | 3242.166305 | 4490.873271 | 1197.944644 |
| | | epsilon | 0.010601542 | 0.166934939 | 0.706444639 |
| GA | MLP-R | activation | 'tanh' | 'logistic' | 'logistic' |
| | | solver | 'lbfgs' | 'lbfgs' | 'lbfgs' |
| | | learning_rate | 'adaptive' | 'constant' | 'adaptive' |
| | | hidden_layer_sizes | 120 | 27 | 196 |
| | SVR | kernel | 'poly' | 'poly' | 'poly' |
| | | C | 851.1898198 | 6628.627568 | 508.4838285 |
| | | epsilon | 0.009785739 | 0.010329552 | 0.001592636 |

**Table 3.** Optimal hyper-parameter configuration selected by different HPO algorithms at different wavelengths.

### Results and discussion

To evaluate the HPO methods, we implemented five fold cross-validation and used RMSE as the performance metric. Additionally, we measured CT as a model efficiency metric. CT is the total time required to complete an HPO process. We specify the same hyper-parameter configuration space (Table 2) for all HPO methods to fairly compare GS, RS and GA. The optimal hyper-parameter configuration (Table 3) was determined by each of the HPO methods based on the lowest RMSE for all three wavelengths.

We tuned the models on a machine with an 8 Core i7-9700K processor and 16 gigabytes (GB) of memory. We used Python 3.5, multiple open-source Python libraries, and open-source Python frameworks, including sklearn[34]. Figure 10 shows that for both MLP-R and SVR, RS produces much faster results than GS while maintaining lower RMSE for the same search space size. In general, GA offers lower RMSE for both models but has a higher CT compared to GS and RS in all three wavelengths. Overall, MLP-R outperforms SVR in terms of performance. However, we achieved better efficiency with SVR in our dataset.

| ML model | HPO method | Error ($\mu \pm \sigma$) at 493.4 nm | Error ($\mu \pm \sigma$) at 821.4 nm | Error ($\mu \pm \sigma$) at 868.1 nm |
|---|---|---|---|---|
| MLP-R | Default | 141.53 ± 84.52 | 183.42 ± 99.15 | 139.88 ± 101.50 |
| | GS | 76.37 ± 61.51 | 132.01 ± 75.05 | 82.19 ± 79.53 |
| | RS | 65.21 ± 59.91 | 113.95 ± 59.28 | 73.88 ± 48.47 |
| | GA | 51.11 ± 43.33 | 90.34 ± 43.02 | 63.80 ± 48.32 |
| SVR | Default | 171.14 ± 99.96 | 142.97 ± 78.00 | 121.23 ± 86.12 |
| | GS | 80.99 ± 53.78 | 105.32 ± 70.34 | 75.97 ± 79.04 |
| | RS | 48.47 ± 49.38 | 73.21 ± 66.30 | 63.68 ± 51.21 |
| | GA | 44.46 ± 33.60 | 69.32 ± 48.12 | 49.54 ± 37.85 |

**Table 4.** Estimation error for predicting soil TN.

We introduced the machine learning approach to estimate the TN of soil using NDVI and multispectral characteristics (R, NIR and G) of the images. We also consider the environmental factors such as air temperature and RH. The performance of MLP-R and SVR models were tested on a fixed configuration space for the hyper-parameters under various hyper-parameter optimization techniques at three different wavelengths (Table 4). For both MLP-R and SVR, the default HP configuration do not yield the lowest RMSE, this demonstrates the significance of utilizing HPO. From Table 4, the estimation error of predicting soil TN is lowest in GA compared to GS and RS for both MLP-R and SVR, where $\mu$ is the mean and $\sigma$ is the standard deviation. While training the models, we split our dataset into train and test for all three wavelengths individually, where we use 80% of the data for training and 20% for testing.

The UMS framework can be used to estimate the total nitrogen in soil. However, depending on the types of soil and crops, the model needs to be re-calibrated. More specifically, the actual TN of soil should be obtained from the subset of the samples to calibrate the N spectrum's intensity after determining the N lines using LIBS. Furthermore, N lines that fall around the 500 nm region should be avoided in sea sand due to the interferences with Titanium (Ti) lines[23].

## Conclusions

In this paper, we have demonstrated the ability of a UAV-based multispectral sensing solution to estimate soil total nitrogen. Specifically, we implemented two machine learning models *multilayer perceptron regression* and *support vector regression* to predict soil total nitrogen using a suite of data classes including UAV-based imaging data in red, near infrared, and green spectral bands, normalized difference vegetation indices (computed using the multispectral images), air temperature, and relative humidity. We performed hyperparameter optimization methods to tune the models for prediction performance. Overall, our numerical studies confirm that our machine learning-based predictive models can estimate total nitrogen of the soil with a root mean square percent error (RMSPE) of 10.8%.

## Data availability

The source code, and the training data can be found here, https://git.io/JOaqK.

### Acknowledgements

This work was supported in part by South Dakota GOED i6 program through the Proof of Concept grant. We thank Dr. Christopher Graham for support in data collection.

### Author contributions

M.A.H. conducted the data collection, experiments and prepared the results. The LIBS experiments were conducted by P.K.D. and M.A.H. Analysis was done by all authors. The manuscript was prepared by M.A.H. and S.R. S. R. served as the principal investigator for this project. All authors reviewed the manuscript.

### Competing interests

The authors declare no competing interests.

### Additional information

**Correspondence** and requests for materials should be addressed to S.R.

**Reprints and permissions information** is available at www.nature.com/reprints.

**Publisher's note** Springer Nature remains neutral with regard to jurisdictional claims in published maps and institutional affiliations.